\documentclass[a4paper,nofootinbib]{revtex4}
\usepackage{graphicx}
\usepackage{amsmath}
\usepackage{amsthm}
\usepackage{amsfonts}
\usepackage{amssymb}
\usepackage{dsfont}
\usepackage{mathrsfs}
\usepackage{graphics}
\usepackage{epstopdf}
\newcommand{\ie}{{\emph{i.e.}}~}
\newcommand{\ket}[2]{|#1 \rangle_{#2}}
\newcommand{\bra}[2]{ \phantom{}_{#2}\langle#1 |}

\newcommand{\lie}{{\pounds}}

\theoremstyle{plain}

\newcommand{\hodge}{\ast}

\newcommand{\dext}{\mathrm{d}_{\mathrm{ext}}}

\newcommand{\GG}{\mathscr{G}}

\newcommand{\eff}{{\mathrm{eff}}}
\newcommand{\ent}{{\mathrm{ent}}}
\newcommand{\grav}{{\mathrm{grav}}}
\newcommand{\micro}{{\mathrm{micro}}}
\newcommand{\matt}{{\mathrm{matter}}}
\newcommand{\inte}{{\mathrm{int}}}
\newcommand{\tot}{{\mathrm{tot}}}
\newcommand{\vac}{{\mathrm{vac}}}

\newcommand{\ham}{\hat{H}}

\newcommand{\htot}{\HH_{\mathrm{tot}}}
\newcommand{\hmat}{\HH_{\matt}}

\newcommand{\hgrav}{\HH_\grav}

\newcommand{\hrho}{\hat{\rho}}

\newcommand{\eg}{{\it e.g.~}}
\newcommand{\hpsi}{\hat{\psi}}

\newcommand{\DD}{\mathcal{D}}
\newcommand{\OO}{\mathcal{O}}

\newcommand{\ee}{\mathbf{e}}
\newcommand{\HH}{\mathscr{H}}

\newcommand{\Tr}{\mathrm{Tr}}

\newcommand{\ini}{\mathrm{initial}}
\newcommand{\fin}{\mathrm{final}}

\begin{document}
\begin{flushright}
AEI-2012-192
\end{flushright}

\title{
Horizon thermodynamics and composite metrics
}

\author{Lorenzo Sindoni}

\address{Max Planck Institute for Gravitational Physics (Albert Einstein Institute), \\Am M\"uhlenberg 1, 14476 Potsdam, Germany}

\email{sindoni@aei.mpg.de}

\begin{abstract}
We examine the conditions under which the thermodynamic behaviour of gravity can be explained within an emergent gravity scenario, where the metric is defined as a composite operator. We show
that due to the availability of a boundary of a boundary principle for the quantum effective action, Clausius-like relations can always be constructed. Hence, any true explanation of the thermodynamic nature of the metric tensor has to be referred to an equilibration process, associated to the presence of an H-theorem, possibly driven by decoherence induced by the pregeometric degrees of freedom, and their entanglement with the geometric ones.
\end{abstract}

\maketitle

\section{Introduction}

In this work we explicitly examine the origin of thermodynamical behavior of horizons,
starting from a microscopic theory and deducing from it a form of emergent gravitational theory. For this we will use a toy model, 
but only to fix the ideas
with explicit calculations. We will show that a key and often overlooked
ingredient is the process
of equilibration 
of the microscopic degrees of freedom leading to decoherence of the degrees of freedom encoded 
into the metric tensor. 
Whether this process takes place or not is ultimately a property
of the dynamics of the microscopic theory, and we argue that it is only within a microscopic 
theory addressing this specific point that the clarification 
of the relationship between gravitational theories and thermodynamics can be fully understood.

The possibility that the gravitational interaction, as described by classical general relativity (GR),
might be after all of thermodynamic/hydrodynamic nature is by now supported by a large amount
of evidences (for a discussion see, for instance, \cite{Padmanabhan:2012qz,Padmanabhan:2009vy} and references therein).
One of the most striking results is that Einstein's equations can in fact be derived starting from a
thermodynamic standpoint: once the equivalence principle is advocated (in the form of the existence of a Lorentzian metric to which matter fields are universally coupled), in each sufficiently small
neighborhood of the spacetime manifold one can construct local acceleration horizons (basically 
local Rindler horizons, null orbits of approximate Killing vector fields), whose dynamics is then assumed to 
be governed, at equilibrium, by a Clausius-like relation $\delta Q = T dS$ \cite{Jacobson:1995ab}. 
This result has been generalized in a number of ways to a number of different situations,
see for instance \cite{Eling:2006aw, Brustein:2009hy,Chirco:2009dc,Guedens:2011dy} and references therein.

In these derivations, after a suitable identification of the entropy functional
associated to the local Rindler horizon (\eg the area of its spatial cross section), the temperature
as the Unruh temperature of a uniformly accelerated observer (associated to the given null surface)
and the heat flow $\delta Q$ as the flux of the conserved stress energy tensor across the horizon,
one can obtain Einstein's equations as equations of state.
In fact, a more careful analysis \cite{Chirco:2009dc} shows that it is necessary to extend the treatment to the case of nonequilibrium thermodynamics (introducing appropriate internal entropy production terms) in order to properly take into account all the contributions appearing 
into the Clausius-like relation. 

Despite the proposal to use concepts like entanglement entropy of the vacuum to
explain the appearance of such a behavior (see \cite{Jacobson:2012yt} for a very recent discussion), it is not yet 
clear whether this is the actual explanation of the correspondence
with thermodynamics, and whether this correspondence is accidental or substantial. 
Indeed, from a pregeometric point of view, it is not obvious why the notion of causality (which is at the basis
of the entanglement entropy arguments) should in any way inform the emergent dynamics, but, most importantly, the 
thermodynamics emerging from the microscopic degrees of freedom. Of course entanglement entropy is an important piece
of the puzzle, and it is certainly correlated with it, but since it pertains to the semiclassical limit of the theory it is not completely obvious that it
is ultimately the origin of the peculiar thermodynamic behavior of gravity.

To settle this issue, however, it is unavoidable
to have at least a qualitative picture of what could be the layer below GR: if we want to explain
GR as thermodynamics (and perhaps get some glimpses of a more ``fundamental" level) we really 
need to have a picture of it in terms of the statistical mechanics of some (yet unknown)
fundamental degrees of freedom. This is a necessary step if we want to give further support to
results on the counting of microscopic d.o.f., especially when stated in the form of
equipartition conditions \cite{Padmanabhan:2009kr,Padmanabhan:2010xh,Padmanabhan:2012bs}.
 
The objective of this contribution is to discuss, at least in a special case, the way in which this
thermodynamical behavior arises, limiting the analysis to a class of models proposed to include the gravitational field as a composite field 
of some other microscopic degrees of freedom. Then, we will try to 
elucidate what are the necessary ingredients needed to establish the macroscopic thermodynamical
behavior from this specific microscopic system. While some ideas are certainly not new, the 
objective is to make the results as explicit as possible in a concrete case. The aim is understand 
what are the key points that have to be clarified in a scenario in which
gravity is seen as an emergent phenomenon \cite{review}, irrespective of the particular
microscopic model adopted.

While the class of models that we will be using does not have a rigorous implementation (essentially for a nonrenormalizability problem), we might still try to use it as a sort of paradigm, a toy model
from which to extract lessons that might be applicable in a more general context,
possibly including a definitive (still unknown in its final form) theory for quantum or emergent gravity.
The great advantage of doing so is that we will be able to derive explicit results using standard
techniques of quantum field theories.

The result is that the quantum effective action for the composite metric exhibits thermodynamical
aspects even in the ``deep quantum regime" where the metric fluctuations can be large, but most importantly where the metric does not exhaust the specification of the macroscopic state.
The aspect that is overlooked, indeed, is the problem of equilibration of the microscopic degrees of
freedom. 

The concepts of decoherence have been already related to the problems of the semiclassical limit of
gravity (see, for instance, \cite{Padmanabhan:1989rm,KieferJoos,Kiefer:1997hv,Garay:1999cy,Mavromatos:2004sz,Gambini:2006yj}), and in particular the role of unobserved
modes in producing an effective decoherence of the wavefunction for geometry. However, a genuinely pregeometric
scenario is not explored, so far. We try to go
beyond this state of affairs, including in the game truly microscopic modes that are, in a sense,
``atoms'' of spacetime and matter fields and show that these microscopic states are really
necessary to show that the thermodynamic behavior of geometrodynamics is not an accident.

Equilibration, in the picture that we are going to examine, turns out to 
be the relevant ingredient that might lead to the explanation of the various thermodynamical
reasonings attached to gravity, as the generalized second law 
\cite{Bekenstein:1974ax}. (For a related discussion, but with a different perspective on the role of 
microscopic degrees of freedom, equilibrium and entropic properties of gravity see also \cite{Kay:1998vv,Kay:1998cj}.)
In extreme summary, the metric will acquire a thermodynamical interpretation only when it can be treated as
a thermodynamic state function, \ie in an equilibrium configuration, by definition, and this
in turns implies the existence of constraints on the microscopic dynamics that has to lead
to equilibration with respect to metric variables. 

The plan of the paper is as follows. In the next section we describe in details the model, the effective action
for the composite metric and show how it supports naturally a thermodynamic interpretation even without a 
restriction to a semiclassical or equilibrium configurations. In section III we will then discuss in detail
the role of decoherence in the appearance of an equilibrium configuration at the macroscopic level,
highlighting the role of entanglement with the microscopic degrees of freedom that are integrated away.
Finally, in the concluding section, we summarize the results and connect them with other approaches
to quantum gravity that try to address the emergence of a continuum semiclassical spacetime.

\section{From Pregeometry to Thermodynamics}
The models that we are considering have been called ``pregeometry'' \cite{Akama:1977hr,Akama:1978pg,Akama:1977nw,Akama:1981dk,Terazawa:1991yj,Amati:1981tu,Amati:1981rf}, even though they are not as pregeometric
as they pretend to be. We briefly sketch the idea underlying the models, referring to the literature for a detailed discussion (see also \cite{Hebecker:2003iw,Wetterich:2003wr,Wetterich:2011yf,Wetterich:2012ps,Wetterich:2012ag,Wetterich:2012np,Wetterich:2011xv,Sexty:2012vr} for very recent developments).
Reduced to the bone, they are the quantization of the Polyakov action of a 
bosonic or fermionic membrane of assigned dimension (for us it will be four) embedded in a flat (pseudo-)Riemannian manifold of equal or higher dimension. The general idea is that the metric tensor as we know it is just a sort of meson field, \ie a composite operator. We will work within a field theory model, and we will denote as $\hpsi_{A}$ ($A=1,...N$) the field operators associated to the microscopic degrees of freedom, where $A$ is a label denoting collectively possible further fine grained details of the field (internal symmetries) 

In the bosonic case (the Fermionic one leading essentially to the same conclusions, within a first order formalism), 
the classical metric will be given, formally\footnote{The metric tensor is not a gauge invariant
quantity, under the action of diffeomorphisms. Hence, we should refer to appropriate gauge invariant
quantities, \ie proper observables. With this caveat in mind, we will keep using the metric tensor
as the vacuum expectation value of an operator.}, by the expectation value
\begin{equation}
g_{\mu\nu}(\Psi) = \bra{\Psi}{} \hat{\mathcal{O}}_{\mu\nu}(\hpsi) \ket{\Psi}{} =
\bra{\Psi}{}\eta^{AB}\partial_{\mu}\hpsi_{A}\partial_{\nu}\hpsi_{B}) \ket{\Psi}{}
\label{compmetric}.
\end{equation}
It is clear, from this very definition, that several distinct microscopic configurations might lead to the same macroscopic metric, given the presence of a global $SO(N-1,1)$ invariance.
Indeed, if we consider a field configuration $\phi^A$,
\begin{equation}
g_{\mu\nu}(\Lambda \phi) = \eta^{AB}\Lambda_{A}^{C}\Lambda_{B}^{D}\partial_{\mu}\psi_{C}\partial_{\nu}\psi_{D} = g_{\mu\nu}(\phi)
.
\end{equation}
This is of course a special case of a possibly more general situation.
The partition function for the system is (rather formally) defined to be
\begin{equation}
Z=\int \DD \phi \DD \gamma \DD\psi
\exp\left( -
 \int d^4x \left\{ (\det\phi_{\mu\nu})^{1/2} + \sqrt{\gamma}\gamma^{\mu\nu}(\phi_{\mu\nu}-
 \eta^{AB}\partial_{\mu}\psi_{A}\partial_{\nu}\psi_{B}) \right\} \right),
\end{equation}
where $\phi_{\mu\nu}$ is an auxiliary field encoding the composite metric,
$\eta^{AB}$ is the (inverse) metric tensor of the embedding flat space (or spacetime in case
of Lorentzian signature), and $\gamma^{\mu\nu}$ are Lagrange multipliers to enforce the constraint:
\begin{equation}
\phi_{\mu\nu} = \eta^{AB}\partial_{\mu}\psi_{A}\partial_{\nu}\psi_{B}.
\end{equation}
Matter fields (singlets with respect to the global $SO(N-1,1)$ symmetry) can be introduced, and coupled to the composite metric without difficulty,
technically. From a conceptual point of view, however, a question naturally arises: why matter fields should be coupled to one or another choice of possible metric tensors?
One possibility is that matter fields are excitations of the underlying microscopic theory, and the coupling is unavoidable.
Using an analogy, one might argue that, similarly to what happens in condensed matter analogues with the coupling
of phonons to the acoustic metric \cite{Barcelo:2005fc},
matter fields are excitations over (or of) the background, which is determining the metric at the same
time.
In this case, the coupling to the metric is essentially forced to be the one we are used to. 

We can introduce a generating functional for the metric as a composite operator:
\begin{equation}
Z[K^{\mu\nu}] = \int \DD \phi \DD \gamma \DD\psi \int d^4x \left\{ (\det\phi_{\mu\nu})^{1/2} + \sqrt{\gamma}\gamma^{\mu\nu}(\phi_{\mu\nu}-\eta^{AB}\partial_{\mu}\psi_{A}\partial_{\nu}\psi_{B}) 
+ K^{\mu\nu}\eta^{AB} \partial_{\mu}\psi_{A}\partial_{\nu}\psi_{B}
\right\},
\end{equation}
such that:
\begin{equation}
g_{\mu\nu} = \left.\frac{1}{Z} \frac{\delta Z[K]}{\delta K^{\mu\nu}}\right|_{K=0}.
\end{equation}
Actually, in order to deal with physical questions, one can evaluate transition amplitudes with
assigned values of the fields $\psi_{A}$ at the two boundaries. The treatment does not change
much, and the resulting effective action depends also on the choice of boundary states, as
expected.

Using standard techniques \cite{Cornwall}, we can define a quantum effective action for this composite operator,
$
\Gamma_{\vac}[g_{\mu\nu}],
$
which is a functional for a classical field that encodes all the quantum effects.
Indeed, we are going to use this object and we will manipulate the metric tensor as a classical field,
even without asking for the semiclassical limit to be enforced, \ie without requiring that the quantum
fluctuations around the classical field are small (to do this we would need to analyze the
quantum effective action for the composite operator associated to the fluctuations of the metric,
\eg for a correlator like $\langle g_{\mu\nu}(x)g_{\rho\sigma}(y)\rangle = D_{\mu\nu\rho\sigma}(x,y)$).

Given the quadratic dependence of the bare action of the fields $\psi$, we can easily integrate them out to get an explicit effective action for the composite metric tensor. 
Standard calculations of quantum field theory in curved backgrounds show that the effective action
for the composite metric
will indeed be given in terms of the divergent contributions (which are cutoff regulated) of the
heat kernel expansion of the one loop effective action, and will have the form of integrals of curvature invariants determined by the HaMiDeW coefficients. 
As it has been noticed very soon, these models do not offer a satisfactory realization of Sakharov's 
idea of induced gravity \cite{Sakharov,VisserSakharov} as the dependence on the cutoff needed
for the regularization of the partition function cannot be hidden via a renormalization
procedure, leading to a non-renormalizability problem.

By construction, in such a model we obtain for the composite metric\footnote{We stress it again, it is a classical field, associated to a quantum state.}  equations of motion having the general form
\begin{equation}
\mathscr{G}_{\mu\nu}(g_{\alpha\beta}) = \kappa T_{\mu\nu},
\end{equation}
with a conserved stress energy tensor $T_{\mu\nu}$ for the matter fields, $\kappa$ a coupling constant, and $\GG$ a covariantly
conserved rank two tensor obtained from the effective action of the composite field $g_{\mu\nu}$.
As said, this tensor will depend on the boundary conditions imposed on the functional integral (or, better, the boundary states). 
We will denote this dependence with a (collective) label $k$, reminiscent of a renormalization group flow parameter.

The thermodynamical behavior of such models can be easily inferred directly by the Noether charge method \cite{Wald:1993nt,Iyer:1994ys,Iyer:1995kg}. However, here we will follow a connected but slightly different approach which
will in turn lead to a more transparent explanation of some of the terms that will appear, interpreted
as non-equilibrium internal entropy production terms, as well as the general structures at work, in order to be able to treat more general models.

Assuming that the metric is at least of class $C^3$, we construct the vector valued 3-forms:
\begin{equation}
\GG^{(k)\mu}_{\phantom{(k)}\nu}\ee_{\mu}\hodge dx^{\nu}, \qquad T^{\mu}_{\nu}\ee_{\mu}\hodge dx^{\nu},
\end{equation}
where $\hodge$ is the Hodge star associated to the metric $g_{\mu\nu}$. With these objects,
the quantum equations of motion can be put in a form that satisfies Wheeler's boundary of a boundary principle \cite{Kheyfets:1986na,Kheyfets:1986tk,Kheyfets:1991at}.
Given that we are dealing with a (pseudo-)Riemannian manifold, around each point $P$ one can 
construct an approximate Killing vector field $\xi^{\mu}$,
\begin{equation}
\lie_{\xi}g_{\mu\nu} = l_{\mu\nu},\qquad |l_{\mu\nu}|\ll \ell^{-1},
\end{equation}
which is generating what appear to be, locally, a piece of a Rindler horizon (here $\ell$ is
the typical scale we are interested in, possibly related to the curvature). Thanks to these 
structures, one can easily look at the equations integrated on a portion $H$ of this Rindler horizon
between an initial and a final slice, $\sigma_{\ini}$ and $\sigma_{\fin}$ (no other boundaries are considered). First, one has to project the vector valued forms onto
the generator of the null surface:
\begin{equation}
\GG^{(k)\mu}_{\phantom{(k)}\nu}\xi_{\mu}\hodge dx^{\nu}, \qquad T^{\mu}_{\nu}\xi_{\mu}\hodge dx^{\nu},
\end{equation}
and then one can work with an integrated version of the equations of motion over three 
dimensional submanifolds. This construction emphasizes the possibility of writing gravitational
equations in an integral form, rather than in a differential way.
Also, it allows a more clear treatment of the physical processes, since it does not require any expansion
in small parameters to highlight the relevant features.

In particular, one can consider these forms integrated on null surfaces
\begin{equation}
\int_{H}\GG^{(k)\mu}_{\phantom{(k)}\nu}\xi_{\mu}\hodge dx^{\nu} = \kappa \int_{H}T^{\mu}_{\nu}\xi_{\mu}\hodge dx^{\nu}.
\label{eq:integral}
\end{equation}
The RHS is just proportional to the heat flow that enters the thermodynamic reasoning of 
\cite{Jacobson:1995ab}.
The analysis of the LHS should lead to the identification of the entropy function, if any,
as it is coming from the microscopic theory. First, let us notice that (provided that $k$ can be considered constant on H) it is the integral of a 3-form,
and as such it can be decomposed (via the Hodge decomposition theorem) into
an exact, coexact and harmonic components
\begin{equation}
\GG_{(k)}(\xi)=\GG^{(k)\mu}_{\phantom{(k)}\nu}\xi_{\mu}\hodge dx^{\nu}  = \dext {\alpha} + \delta \beta + \gamma,
\end{equation}
where $\delta$ is the codifferential, $\alpha$ is a two form, $\beta$ a 4-form, and $\gamma$ a harmonic 3-form
(associated to global properties of the manifold).
Second, due to the covariant conservation of the tensor $\GG^{\mu\nu}$, the coexact part of this form is zero whenever the vector field $\xi^{\mu}$ is a Killing vector field, and the heat flow is exactly measured in terms of the variation of certain quantities evaluated at the initial and final slices of the local horizon (if the harmonic component vanishes, otherwise an additional
non-equilibrium term has to be included). 

The entropy function to be inserted into a Clausius-like equation, then, comes from the exact part of the 
3-form $\GG_{(k)}(\xi)$.
\begin{equation}
\int_{\partial H}\alpha_{k}(\xi) + \int_{H} (\delta \beta_{(k)}(\xi) +\gamma_{(k)}(\xi)) = \kappa \delta Q,
\end{equation}
where $\partial H=\sigma_{1} \cup \sigma_{2}$ (with the appropriate orientations) is the boundary of the portion of horizon chosen (which we are implicitly assuming to have the topology of a cylinder, without loss of generality). 
The Clausius-like relation suggests to identify the equilibrium entropy terms as
\begin{equation}
S_{k}(\sigma_{i};\xi) \propto\int_{\sigma_{i}} \alpha_{(k)}(\xi),
\end{equation}
the proportionality factor to be determined once the temperature is identified. 

This derivation shows that these models easily support a (non-equilibrium) 
thermodynamical interpretation and predict (within the limits of
their domain of definition) the various terms appearing in a Clausius-like relation, including the
internal entropy production terms, which are related to the coexact part of $\GG_{(k)}(\xi)$ 
(associated to the fact that $\xi$, the generator of the null surface, is not a Killing vector field)
and to the harmonic component of the curvature's flux 3-form.

This construction has the nice feature of explicitly showing how the equilibrium entropy term arises, and that it is accompanied
by nonequilibrium terms associated not to the failure of the effective action to have a
certain functional form, but rather to the failure of the local horizon to be a Killing horizon.
This whole construction, we stress it again, is ultimately possible because the effective equations for the metric can be put in a form that naturally follows 
from Wheeler's boundary of a boundary principle.
In fact, if we work in $d$ spacetime dimensions, the generalized Cartan moment of rotation $\GG$ will
be a $d-1$ form, and hence the equations of motion can be naturally expressed in terms of relations
between fluxes (of tensors, suitably projected) across the boundaries of $d$ dimensional domains of the spacetime
manifold, in particular across causal boundaries given by horizons.

Nonetheless, it is clear that this relation between the effective dynamics of the composite metric
tensor and non-equilibrium thermodynamics is rather formal, at least at this stage. Indeed, the effective action 
for the metric crucially
depends on the assigned boundary data, that are typically given in terms of the microscopic
field configurations, which are unobservable by matter fields coupled only to the metric tensor. 
Also, it is a property that follows immediately from the shape of the equations of motion for the
classical metric tensor (essentially, diffeomorphism invariance), whether the physical state is a semiclassical one or not, \ie whether the
quantum fluctuations of the metric tensor are small or not.

In fact, the expectation value of the (spatial part of the) metric tensor (or the extrinsic curvature, if we use it instead of the metric) at the initial and final slices is not the only quantity that is fixed with the specification of the boundary data via the $\psi_{A}$ fields, but also the fluctuations and all the possible composite fields that is possible to construct from the
microscopic degrees of freedom. Even though not all of them will turn out to be independent, at
the macroscopic level the state is specified by more than the metric tensor on the boundary,
and this will become manifest from the fact that states with the same boundary metric data will evolve
differently since the (expectation value of the) boundary metric does not exhaust the information
encoded in the label $k$ introduced before to collectively denote all the microscopic data.

\begin{figure}
\begin{center}
\includegraphics{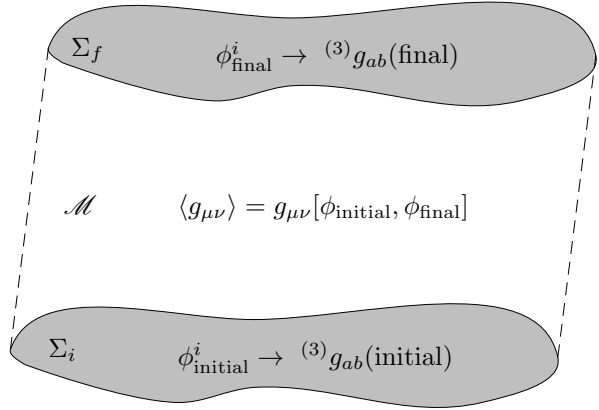}
\end{center}
\caption{A pictorial representation of the data assignment for the transition amplitudes.}
\label{fig:transition}
\end{figure}

An easy way to see this is the following. Consider the transition amplitude of the fundamental theory between two field configurations defined on the two disconnected boundaries, as depicted in Fig. \ref{fig:transition}.
The transition amplitude and the metric tensor in the bulk are functionals of the initial and final 
{\emph{microscopic}} data,
\begin{equation}
\mathcal{A}[\psi_{\ini},\psi_{\fin}], \qquad g_{\mu\nu}[\psi_{\ini},\psi_{\fin}].
\end{equation}

According to the definition \eqref{compmetric}, specialized to the 3D components, the metric
tensors of a given configuration $\psi_{\fin}$ and of a modified one $\Lambda \psi_{\fin}$ do coincide, but the amplitude, in principle, is changed, 
given that the symmetry $SO(N-1,1)$ that we are using is global, and not local.
This happens unless the $SO(N-1,1)$ symmetry from global becomes local, at least
approximately, at the dynamical level. 

An extreme case of this complication can be seen easily from the following argument. 
Due to the presence of a global symmetry, the associated superselection rule will tell us
that configurations with different charges, albeit possessing the same boundary metrics, will
lead to zero amplitudes. 
Of course, a model that is invariant under local $SO(N-1,1)$ at
the level of the fundamental action would never have this problem. However, that class of models
would require at least a connection structure to be introduced from the very beginning, 
a thing that is clearly against the rules of the game, since it would be a specification of
a background geometric structure.

Of course, in the original proposal for the model, this puzzling connection with thermodynamics is related to the fact that the metric, as treated here, is not really
a thermodynamical variable, ideally uniquely specifying the macroscopic state of the full
set of invisible microscopic degrees of freedom: a microscopic equilibrium regime has to be established, so that the specification of the spatial metric (and/or the extrinsic curvature) on the boundaries uniquely determines
the (possibly mixed) macroscopic state of the system, in such a way that the equations of motion
for the metric tensor can really be interpreted as a form of thermodynamics.

\section{Equilibration}

The notion of equilibrium is deeply related to a notion of time, and hence it is a delicate subject
to be discussed in theories with reparametrization invariance, such as general gravitational theories and
the class of models that
we are playing with. Therefore, a deparametrization of the theory in terms of a suitable definition of clock variable has to be performed. We will assume that it is possible to single out a clock, such
that it is possible to define the condition of equilibrium as a condition of stationarity of macroscopic
variables under the evolution of a proper Hamiltonian. We will then
assume that, instead of working with a Wheeler--De Witt equation, we will work with a proper
Schr\"odinger equation for the full density matrix of the deparametrized system,
\begin{equation}
i \hbar \frac{\partial }{\partial t} \hrho = [\hrho,\ham],
\end{equation}
where $t$ is the time associated to the clock and $\ham$ is the resulting Hamiltonian.

The key to equilibration will be some form of effective decoherence of the reduced density matrix
associated to the system containing the composite metric tensor and the matter fields. The ideas
that we are going to discuss here go along the line of the attempts made to establish thermalization
and the rise of the canonical ensemble from a detailed derivation from the principles of quantum mechanics \cite{2010NJPh...12e5027R,2012NJPh...14d3020R,2012PhRvL.108h0402R}.

In order to highlight the line of reasoning in a clear way, we will use tensorial notation. To fix the
notation,
given a basis of a Hilbert space labeled by a certain set of indices $i$, instead of working
with the density matrix as an operator, we will use its matrix coefficients as follows:
\begin{equation}
\hrho = \sum_{i,j} \rho^{i}_{j}\ket{i}{}\bra{j}{};\qquad \rho^{i}_{j} = \bra{i}{}\hrho\ket{j}{}.
\end{equation} 

The full Hilbert
space will be decomposed in the tensor product of three Hilbert spaces, 
\begin{equation}
\htot =  \HH_{\micro}\otimes \hgrav \otimes \hmat,
\end{equation}
where $\HH_{\micro}$ is associated to the invisible microscopic
configurations (for which we will use upper case latin indices $A,B,C$), $\hgrav$ is associated to the degrees of freedom giving the metric (for which we will
use greek indices $\alpha,\beta,\gamma$), and, finally,
$\hmat$ is associated to the matter degrees of freedom used to probe geometry (for which
we will use latin indices $i,j,...$).
\footnote{Given the full density matrix for the full tensor
product of the two systems, A and B,
the reduced density matrices are:
\begin{equation}
\rho^{a}_{b}(\mathcal{A}) 
= 
\rho^{ai}_{bj}\delta^{j}_{i};\qquad \rho^{i}_{j}(\mathcal{B}) = \rho^{ai}_{bj}\delta^{b}_{a}.
\end{equation}
We then parametrize the full density matrix as follows:
\begin{equation}
\rho^{ai}_{bj} = \rho^{a}_{b}(\mathcal{A})\rho^{i}_{j}(\mathcal{B}) + \rho^{ai}_{bj}(\ent),
\end{equation}
where $\rho(\ent)$ encodes the entanglement between the two subsystems. It follows immediately that this matrix is traceless
in both pairs of indices. }
We now proceed and trace away the microscopic degrees of freedom of our system:
\begin{equation}
\tilde{\rho}^{\alpha i}_{\beta j}=\rho^{A \alpha i}_{B\beta j }\delta^{B}_{A}.
\end{equation}
The equation of motion for the reduced density matrix is easily obtained from the fundamental
Schr\"odinger equation\footnote{We do not write it in the form of a Lindblad equation \cite{Lindblad:1975ef}
in order to make more evident the contributions coming from the entanglement terms with
the microscopic degrees of freedom.}:
\begin{equation}
i\hbar \frac{\partial}{\partial t}\tilde{\rho}^{\alpha i}_{\beta j} 
= [H_{\eff},\tilde{\rho}]^{\alpha i}_{\beta j} + i \hbar \Xi^{\alpha i}_{\beta j},
\end{equation}
where 
\begin{equation}
(H_{\eff})^{\alpha i}_{\beta j} = H^{A\alpha i}_{C\beta j}\rho^{C}_{A}(\micro)
\end{equation}
is a (time dependent) effective Hamiltonian including the direct effects of the microscopic degrees
of freedom on the dynamics (the one appearing in a renormalization group transformation of the Hamiltonian after
a block transformation, for instance), and 
\begin{equation}
i\hbar \Xi^{\alpha i}_{\beta j} = [H,\rho(\ent)]^{A \alpha i}_{A\beta j}
\label{eq:entop}
\end{equation}
is a (time dependent) term that describes the loss of coherence due to the tracing over the microscopic degrees of
freedom.

One can further decompose $\tilde{\rho}$ into the gravity and matter subspaces, working along
the same lines. Let us use the simplifying assumption that matter fields represent small
disturbances on the system, and that the entanglement between gravitational degrees
of freedom and matter degrees of freedom is negligible. The effective equation for gravitational
the degrees of freedom is, of course,
\begin{equation}\label{eq:liouvmod}
i\hbar \frac{\partial}{\partial t}{\rho}^{\alpha }_{\beta }(\grav) = [H_{\eff}(\grav),\rho(\grav)]^{\alpha}_{\beta} + i\hbar \Xi^{\alpha}_{\beta}(\grav),
\end{equation}
while the one for matter fields is:
\begin{equation}
i\hbar \frac{\partial}{\partial t}{\rho}^{i}_{j }(\matt) = [H_{\eff}(\matt),\rho(\grav)]^{i}_{j} + i\hbar \Xi^{i}_{j}(\matt).
\end{equation}
As expected, decoherence is appearing in both gravity and matter sectors. However,
the strength of the decoherence effects might be different in the two cases, since it
ultimately depends on the coupling with the microscopic degrees of freedom, to
which matter degrees of freedom are decoupled, as we have assumed\footnote{Obviously, decoupling does not imply absence of entanglement.}. 

In particular, in order to promote the metric degrees of freedom to thermodynamic degrees
of freedom, even in a regime where semiclassical gravity applies (and decoherence in quantum
matter fields is negligible), we need that $\Xi(\grav)$ has a large effect on the matrix $\rho(\grav)$, while decoherence on the matter sector should act more weakly.
This is reasonable in a scenario like pregeometry.

All this derivation is totally general, and it is at this point that we need to discuss equilibration.
Equation \eqref{eq:liouvmod} has not exactly the form of a Liouville equation, given the
presence of the term generated from the entanglement with the microscopic degrees of freedom.
If we move to an interaction representation
\begin{equation}
\hrho_{\inte}(\grav) = \hat{U}_{\eff}^{\dagger} \hrho(\grav) \hat{U}_{\eff},
\qquad \hat{U}_{\eff} = \mathrm{T}\exp\left( -\frac{i}{\hbar} \int dt \hat{H}_{\eff}(\grav) \right),
\end{equation}
the equation of motion reads:
\begin{equation}
\frac{\partial}{\partial t}{\hrho_{\inte}}(\grav) =  \hat{\Xi}_{\inte}(\grav),
\end{equation}
and its solution is evident, at least formally (given that the source term is actually not independent
from the reduced density matrix).
Just to fix the ideas, if we can consider $\hat{\Xi}_{\inte}(\grav)$ essentially as a random 
noise in a sort of Markovian approximation of the full dynamics, we might expect that the latter 
approaches an asymptotic equilibrium configuration, $\hat{\rho}_{0}(\grav)$:
\begin{equation}
\hrho_{\inte}(\grav) = \hrho_{\inte}(\grav)(t=0) + \int^t dt'  \hat{\Xi}_{\inte}(\grav) \rightarrow 
\hrho_{0}(\grav)(t),
\end{equation}
where this last equation has to be intended in a coarse grained sense, \ie that the late time
evolution of the density matrix mimics the equilibrium one only for a certain subclass of
observables. In order to be specific about the time scales and the kind of observables
that have to be considered, a detailed analysis of the dynamics is unavoidable.
It is not unreasonable to expect that the equilibration time has to be of the order of the Planck time,
for gravity. 

Whether equilibration takes place or not strongly depends on the underlying dynamics, which influences
also the kind of information about the initial state that is lost and the one that is preserved. 
Most significantly, if we want that the equilibration preserves the initial assignment of the spatial
metric tensor (and, analogously, for the extrinsic curvature), we need
\begin{equation}
\Delta g_{ij}= \Delta t_{eq}\frac{\partial g_{ij}}{\partial t}  =\frac{\partial }{\partial t} \Tr\left( \hat{ \rho}(\grav)  \hat{g}_{ij} \right) =
\Delta t_{eq}\left[ \Tr\left( \hat{ \Xi}(\grav)  \hat{g}_{ij} \right)
+
\Tr\left( { 
\hrho}(\grav) \frac{\partial }{\partial t} \hat{g}_{ij} \right)\right]
\approx 0,
\end{equation}
a first family of constraints imposed upon the microscopic dynamics, ensuring that  the evolution of the metric tensor is significant only for macroscopic processes, seen as a sequence of
infinitesimal transformations between equilibrium states, as it should be.
It is only when this condition is met that the metric can be treated as a true thermodynamic
variable, given that it would (at least partially) specify the equilibrium density matrix at fixed macroscopic
conditions. To complete the plan, one has to show how many fields, besides the metric (and
the extrinsic curvature), are needed to completely specify the equilibrium state.

It is interesting to remark that decoherence is advocated, here, not necessarily to obtain a semiclassical limit, but rather an equilibrium configuration, and in this we slightly depart from
previous considerations \cite{Padmanabhan:1989rm,KieferJoos,Kiefer:1997hv,Garay:1999cy,Mavromatos:2004sz,Gambini:2006yj}: the semiclassical limit might require an additional step.
Notice also that deviations from equilibrium thermodynamics (\ie the balance between the various terms entering a Clausius-like relation) originated by incomplete equilibration are
different, in principle, from the non equilibrium term associated to the lack of Killing vector fields
as discussed at the beginning, and might be thought as a significant running of the collective
index $k$ in the regions of integration for \eqref{eq:integral}.

This very general discussion already shows that, in order to provide a convincing explanation
of the thermodynamic behavior of gravity in a 
scenario where gravity is emergent, we have to compute much more than the effective action
of the metric as a composite operator. In fact, one should prove a form of H-theorem, which
would select a preferred class of density matrices as equilibrium states, parametrized by the metric
and derived quantities. This is far from
obvious and exceeds the limits of the present discussion.

Establishing such a theorem might be crucial in the completion of the program as it
should also provide a microscopic interpretation of the generalized second law.
Let us consider the von Neumann entropy of the $\tilde{\rho}$ density matrix, neglecting possible
matter/gravity entanglement terms\footnote{These can be included, but do not change the discussion, if not for the addition of more terms.}. If we define the von Neumann entropy of the
macroscopic modes (metric and matter fields) as
\begin{equation}
S_{\tot}=S(\tilde{\rho}) = -\Tr\left[\log(\hat{\tilde{\rho}})\hat{\tilde{\rho}}\right] = S(\hrho(\grav)) + S(\hrho(\matt)),
\end{equation}
then its time derivative
\begin{equation}
\frac{\partial}{\partial t}S_{\tot} =-\Tr \left( \left[ \log \tilde{\rho} + 1 \right] \frac{\partial}{\partial t}(\hrho(\grav)\otimes\hrho(\matt))\right) = - \Tr \left( (\log \tilde{\rho}) \hat{\Xi}\right)
\end{equation}
involves explicitly the entanglement operator $\hat{\Xi}$ (introduced in \eqref{eq:entop}) encoding the effect of the microscopic
degrees of freedom that have been traced away.
An H-theorem can be established if it is possible to argue that the RHS of the last equation is non-negative, leading then to 
\begin{equation}
\frac{\partial}{\partial t}\left(S(\hrho(\grav)) + S(\hrho(\matt))\right) \geq 0.
\end{equation}

This is not yet the generalized second law, because while we might have argued that the metric
controls directly $\hrho(\grav)$ and hence $S_{\grav}$, we have not proven that a local version of
the von Neumann entropy density flow is given by the integrated version of the equations
of motion of the metric coupled to the conserved stress energy tensor, and so that 
the sum of entropy associated to event horizons and the entropy of matter field in the exterior region is a nondecreasing function of time. This is another step
that requires further work, but these preliminary considerations are encouraging. 

Consider now the case in which the metric is perturbed by the presence of some
sources (see also \cite{Bianchi:2012} for a similar analysis). The system responds by its density matrix, and hence the metric gets perturbed
$
g_{ij} \rightarrow g_{ij} + h_{ij}
$.
Then, at first order in $\delta \rho$
\begin{equation}
\delta S = - \delta \Tr\left( \rho \log\rho\right) =  -  \Tr \left( \log(\rho_{0}) \delta \rho \right).
\end{equation}
Without loss of generality,
\begin{equation}
\rho_0 = \exp(- \OO_0),
\end{equation}
where $\OO_0$ is a suitable operator.
Then
\begin{equation}
\delta\rho_{0} = - \rho_0 \int d^3 x \frac{\delta \OO_0}{\delta g_{ij}(x)} \delta g_{ij}(x)   ;
\qquad
\delta S =  \int d^3 x \left\langle  \OO_0 \frac{\delta  \OO_0}{\delta g_{ij}(x)}  \right\rangle \delta g_{ij}(x).
\end{equation}

The tensor density
\begin{equation}
\mathfrak{S}^{ij}=\left\langle  \OO_0 \frac{\delta  \OO_0}{\delta g_{ij}(x)}  \right\rangle
\end{equation}
plays the role of thermodynamically conjugate variable to the metric tensor, and contains, indeed,
the information about the macroscopic statistical state, as it is specified in $\rho_0$ (or $\OO_0$).
The condition of equilibrium coming from a maximum entropy principle is
\begin{equation}
\mathfrak{S}^{ij} = 0,
\end{equation}
which should imply, for self-consistency, that along a physical trajectory the equations of motion of gravity are satisfied as a consequence of equilibrium/maximization of an entropy function,
with the appropriate
boundary conditions.
The second variation of the operator $\OO$,
\begin{equation}
\mathfrak{T}^{ijhk}(x,y) =  \frac{\delta \mathfrak{S}^{ij}(x)}{\delta g_{hk}(y)},
\end{equation}
is a bi-tensor that will determine the amplitudes of fluctuations of the metric tensor, and will be useful
to discover further details of the microscopic theory, once a framework to include fluctuations
is consistently developed (see, for instance, \cite{Hu:2003qn}).
\section{Discussion and conclusions}

Equilibration and thermalization are concepts that are witnessing a revived interest, in recent years.
In particular, the assumptions and requirements at their foundations are reconsidered in light
of the necessity of solid foundations for statistical mechanics and thermodynamics, particularly
for quantum systems. The transition from quantum to classical gravity is just another example to which
these investigations might be applied.

In extreme summary, the lesson that we can get from the example discussed in this paper is the following. 
Assume that we have a
microscopic model for gravity, such that the metric tensor can be seen as a composite operator.
Assume also that it is possible to prove that the metric tensor possesses a diffeo-invariant quantum effective action. Then it is possible to rewrite the quantum equations of motion in thermodynamical
terms, possibly including nonequilibrium terms. However, this is possible even without explicitly requiring anything about a semiclassical limit: the quantum fluctuations of the metric might still be large (however, there might be
conditions on the size of the curvature).

Notice that the construction makes reference only to the possibility to write the equations
of motion in terms of relations among vector valued differential forms, naturally adapted to
describe the flux of energy and curvature across boundaries of domains of spacetime.
Furthermore, we need that diffeo-invariance is preserved, so that Bianchi identities apply to the
various tensors needed. 
In particular, it is interesting to understand whether a discrete setting like quantum Regge calculus
(or spinfoams, coming from a different perspective)
reproduces, in a coarse grained sense, the same property for the quantum effective action. 
Recent discussions \cite{Bianchi:2012ui,Smolin:2012ys} provide encouraging hints in this direction,
even though one should not be too surprised that discretized versions of gravitational theories 
might already be informed of the boundary of a boundary principle, at the heart of the thermodynamic
reasoning that is commonly considered.

In one way or another, however, this shows that the crucial ingredient in understanding gravity
is not just the computation of an appropriate effective action, but also the understanding of
the onset of the semiclassical limit in terms of an equilibration process, in which the metric
tensor essentially becomes the only macroscopic parameter controlling the full density matrix
of the microscopic degrees of freedom, \ie a thermodynamic state function in the technical sense. 
In order to do that one has to prove an H-theorem, in
some form, which would lead to a class of distinguished equilibrium density matrices. Alternatively, if the construction holds also for discrete models, at least in
a coarse grained sense, then the continuum limit might already be associated to an equilibration
of the density matrix, which would be controlled just by the universality class of the phase
transition associated to the continuum limit.

This procedure is also strongly related to the interpretation of the entropy associated to the gravitational field, 
as it emerges in these models. This entropy, provided that the equilibration
is reached, is associated to the microscopic degrees of freedom which are leading to the macroscopic metric tensor. 
It would be a form of entanglement entropy, but it will not be obviously related to the operation of
tracing over degrees of freedom which are associated to causally disconnected regions of spacetime, 
as it has often been advocated in the past (see \cite{Jacobson:2012yt} for a very recent discussion). 
Rather, it is associated to space
itself (or spacetime in a covariant formalism) when it is seen as a subsystem, a natural perspective
in an emergent gravity scenario.

In concluding this brief discussion on the problem of gravity and thermodynamics, it is interesting to 
confront these ideas to a different point of view which is arising in a different proposal for the realization of gravity as an emergent 
collective phenomenon.
In particular, we want to argue that equilibration is not necessarily the only path to be followed.

Group field theories and tensor models \cite{Gurau:2011xp,Oriti:2011jm} are a class of models designed to
generate combinatorially the path integral for the gravitational field in terms of the Feynman
expansion of the generating functional of certain quantum/statistical systems.
In these models, each Feynman diagram can be associated to a (simplicial) complex, \ie
a piecewise flat manifold of given dimension, eventually with further data associated to the
various structures characterizing it. Ideally, the amplitude of each of these Feynman diagrams
has to be associated to the exponential of the corresponding classical gravitational action 
appropriately discretized on the given complex with the given data.

These models start from discrete building blocks. Hence, to recover
GR the continuum and semiclassical limit has to be properly discussed. The continuum limit has been identified
with phase transition in the GFT/tensor model partition function. However, no conclusive
result on the appearance of a meaningful semiclassical limit is available.

A partial analysis, presented in \cite{emergentGFT} suggests that, if we consider only global information as the total volume of a compact spacelike slice (\eg a scale factor), its expectation
value,
 its dynamics and the associated density matrix (even though the latter was represented by a distribution function on the space of graphs, rather than a proper density matrix) for the microscopic and macroscopic degrees of freedom should depend only on the way in which the phase transition is approached,
\ie on its universality class.
While this is certainly an attractive alternative to equilibration as it is considered here, research in this direction is still in
progress, in particular in order to establish a connection with horizon thermodynamics, that requires
a local reasoning that is so far not understood.
Nonetheless we can say that the analysis of the fluctuations of the metric tensor
 around its expectation value, and their dynamics, can shed new light on the problem of
 emergent gravity and might be used to discriminate between microscopic models, even
 in a cosmological setting \cite{Magueijo:2006fu}.

\acknowledgments{ I would like to thank Goffredo Chirco for extensive discussions on thermodynamics and gravity from the point of view of differential forms and Arnau Riera for many discussions on the problem of thermalization.}

\bibliographystyle{apsrev}
\bibliography{biblio}
\end{document}